\documentclass{JHEP3}
\usepackage{epsfig}
\usepackage{amsmath}
\usepackage{mathrsfs}

\newcommand{\be}{\begin{equation}}
\newcommand{\ee}{\end{equation}}
\newcommand{\ba}{\begin{eqnarray}}
\newcommand{\ea}{\end{eqnarray}}
\newcommand{\ra}{\rangle}
\newcommand{\la}{\langle}
\newcommand{\st}{\scriptstyle}
\newcommand{\sst}{\scriptscriptstyle}

\title{QFT with Twisted Poincar\'e Invariance and  the Moyal Product}

\author{Euihun Joung\\
	Laboratoire APC, Universit\'e Paris VII,
	B\^atiment Condorcet,  75205 Paris Cedex 13, France\\
	joung@apc.univ-paris7.fr}
\author{Jihad Mourad\\
	Laboratoire APC, Universit\'e Paris VII, B\^atiment Condorcet,
	75205 Paris Cedex 13, France\\
	mourad@apc.univ-paris7.fr}

\abstract{We study the consequences of twisting the Poincar\'e
invariance in a quantum field theory. First, we construct 
a Fock space compatible
with the twisting and the corresponding creation and annihilation operators.
Then, we show that a covariant field linear 
in creation and annihilation operators does not exist.
Relaxing the linearity condition, a covariant field can be determined. 
We show that it is related to the 
untwisted field by a unitary transformation and the 
resulting n-point functions coincide with the untwisted ones.
We also show that invariance under the twisted symmetry can be realized
using the covariant field with the usual product or by
a non-covariant field with a Moyal product. The resulting
$S$-matrix elements are shown to coincide with the untwisted ones up
to a momenta dependent phase.}

\begin{document}

\section{Introduction}

Invariance under the Poincar\'e group leads to important 
restrictions on quantum field theory.
In the free linear case, it completely determines the theory \cite{Weinberg:1995mt}.
It is then of great interest to examine the consequences of deforming this invariance.
Many possible deformations have been considered in the past as, for instance,
replacing the Poincar\'e group by
(A)dS group  or deforming the Lie algebra structure.

Here, we will examine the possibility of keeping unchanged the algebra structure of
the Poincar\'e universal enveloping algebra (UEA) 
but deforming its coalgebra structure. The latter is crucial in quantum field theory since
the coproduct determines the transformation laws of a system of several particles.
A simple way to deform the coproduct $\Delta$ can be  obtained from
the usual coproduct $\Delta_0$ and a twist element ${\cal F}$ \cite{Drinfeld:1989st}
as
\be
	\Delta={\cal F}\,\Delta_0\,{\cal F}^{-1}\,,
\ee
the resulting UEA will be called in the following the  
 {twisted} Poincar\'e UEA.\footnote{In fact, a theorem by Drinfeld \cite{Drinfeld:1989st}
shows that all deformations of the coproduct of a semi-simple Lie algebra are of this form.}
A consistent example was recently
 \cite{Chaichian:2004za,Reshetikhin:1990ep}
 considered and is given by
\be
	{\cal F}=e^{\frac i2\theta^{\mu\nu}\,{\cal P}_\mu\otimes{\cal P}_\nu}\,,
	\label{Moyal twist}
\ee
where $\theta^{\mu\nu}$, elements of an anti-symmetric matrix, are the
deformation parameters.
This example arose when considering quantum field theory (QFT)
on the noncommutative space \cite{Madore:2000aq,Douglas:2001ba,Szabo:2006wx} where coordinates satisfy
the commutation relations:
\be
	[\,\hat{x}^\mu\,,\,\hat{x}^\nu\,]=i\,\theta^{\mu\nu}\,.
\ee
It describes the low energy effective theory on D-branes with an external $B$-field.
By the Weyl-Moyal correspondence, the theory can be reformulated
as a field theory on commutative space
but the usual product between two fields being replaced by the Moyal $*$-product,
\footnote{We use a simplified notation $v\theta w$ for $v_\mu \theta^{\mu\nu} w_\nu\,$.}
\be
	(f*_{\sst\theta}g)(x)=e^{\frac i2 \partial_x \theta \partial_y}f(x)\,g(y)\,|_{x=y}
	=m({\cal F}^{-1}\rhd (f\otimes g) )(x)\,,
\ee
where $m(f\otimes g)(x)$ is the usual pointwise product $f(x)g(x)\,$.
The existence of the non-invariant matrix $\theta$ in the Moyal product
makes the theory non-covariant under Lorentz transformations.

It was argued \cite{Chaichian:2004za,Chaichian:2004yh}
(see \cite{Majid:1996kd,Oeckl:2000eg,Grosse:2001pr,Wess:2003da,
Koch:2004ud,Bu:2006ha,Gracia-Bondia:2006yj} for related works)
that even though this noncommutative field theory (NCFT) with the Moyal product
is not covariant under the Poincar\'e UEA,
it is nevertheless covariant under the twisted Poincar\'e 
UEA with $\cal F$ given by eq.(\ref{Moyal twist}).
More precisely, it was noticed that the action of an element ${\cal X}$ of the
twisted Poincar\'e UEA on the algebra of functions verifies:
\be
	{\cal X}\rhd (f*_{\sst\theta}g)=
	({\cal X}_{(1)}\rhd f)*_{\sst\theta}({\cal X}_{(2)}\rhd g)\,,
	\qquad
	\Delta({\cal X})={\cal X}_{(1)}\otimes {\cal X}_{(2)}\,.
\ee

On the other hand it was also advocated 
that in order to implement this new covariance,
one should deform also the commutation relations of
creation and annihilation operators \cite{Balachandran:2005eb}
(see also \cite{Tureanu:2006pb,Zahn:2006wt,Balachandran:2006pi} 
for further discussions) and
in that case, it was concluded that there is no physical difference from the 
commutative QFT \cite{Balachandran:2005pn}. In addition, it was
shown that Wightman functions with the Moyal $*$-product
coincide with the untwisted ones \cite{Fiore:2007vg}.
The field operator used in these works is not covariant under the Poincar\'e transformations.
One of our goals in this letter will be to find a covariant field operator.
This operator allows a transparent and manifest 
implementation of the twisted symmetry
without using the Moyal product. We shall show
that the n-point functions are identical to the untwisted ones 
and that the $S$-matrix elements coincide with the usual ones 
up to a momenta dependent phase.

Our aim is to study the consequences of the covariance  
of the quantum field under the twisted Poincar\'e UEA.
First, in Section 2, we construct
the {twisted} Fock space and the associated creation and 
annihilation operators, $a^\dagger_\theta(p)$ and $a_\theta(p)\,$,
are determined 
 in Section 3. We show that fields linear in
 $a^\dagger_\theta(p)$ and $a_\theta(p)\,$ cannot be covariant.
 In particular, replacing in the usual untwisted field operator $\Phi_0$ 
the creation and annihilation operators by $a^\dagger_\theta(p)$ and $a_\theta(p)\,$,
we get a field $\Phi_{{\rm nc}\,\theta}$ which is not covariant.
This is the field considered in \cite{Balachandran:2005eb,Fiore:2007vg}.
A covariant field operator $\Phi_\theta$ can however be found by relaxing the linearity requirement.
We show that n-point functions are related by the simple relation:
\ba
	&&
	\la\Omega|\,(\Phi_\theta*_{\sst\vartheta}\cdots *_{\sst\vartheta} \Phi_\theta)(x_1,\cdots,x_n)\,
	|\Omega\ra
	\nonumber\\
	&&=
	\la\Omega|\,(\Phi_{{\rm nc}\,\theta}*_{\sst\vartheta-\theta}\cdots *_{\sst\vartheta-\theta}
	\Phi_{{\rm nc}\,\theta})(x_1,\cdots,x_n)\,
	|\Omega\ra
	\nonumber\\
	&&=
	\la\Omega|\,(\Phi_{0}*_{\sst\vartheta}\cdots *_{\sst\vartheta} \Phi_{0})(x_1,\cdots,x_n)\,
	|\Omega\ra\,.
\ea
The covariant n-point functions are the ones with $\vartheta=0\,$.
This shows that the covariant theory based on twisted Poincar\'e and the untwisted one 
have identical correlation functions.
It also shows that twisting the product without
using twisted fields, as is usually the case in NCFT, does not lead to a covariant result.
We also argue that the implementation of the twisted covariance in an interacting theory 
can be realized by an interaction Hamiltonian which is local in the field $\Phi_\theta$
or equivalently non-local in the $\Phi_{\rm nc\theta}$ field operator.
If we insist on having asymptotic states transforming covariantly under
the twisted Poincar\'e symmetry then, as is shown in Section 3,
  the $S$-matrix elements are
identical to the untwisted ones up to a momenta dependent phase.
The Appendix contains the proof of some technical results used in the text.

\section{Twisted Poincar\'e UEA and its representation}

The twisted coproduct yields transformation laws of
multi-particle states under the Poincar\'e group
which are modified with respect to the usual ones.
In particular, the action of the UEA on the tensor product
of identical particles does not commute any more
with the permutation of particles,
leading to a twisted Fock space.
In this section,
 we shall briefly review the twisted Poincar\'e UEA representation 
 and use it  for the construction of the Fock space.

Let ${\cal U}$ be the UEA of the Poincar\'e Lie algebra generated by ${\cal P}_\mu$ and ${\cal M}_{\mu\nu}$.
The one-particle Hilbert space ${\mathscr H}$ of a scalar 
particle corresponds to
the scalar unitary irreducible  representation (UIR). Let 
$U({\cal X})$ be the corresponding 
unitary representation of ${\cal X}\in{\cal U}\,$.
A basis of ${\mathscr H}$ is given  by the momentum eigenstates $|p\ra\,$.

The tensor product representation $U^{(n)}({\cal X})$ on ${\mathscr H}^{\otimes n}$,
the tensor product of n copies of ${\mathscr H}$, is obtained from 
 $U({\cal X})$ and 
the coassociative coproduct $\Delta$ as
\be
	U^{(n)}({\cal X})=U^{\otimes n}(\Delta^{(n)}({\cal X}))=
	U({\cal X}_{(1)})\otimes\cdots\otimes U({\cal X}_{(n)})\,,
	\label{tpr}
\ee
where the n-coproduct 
$\Delta^{(n)}({\cal X})={\cal X}_{(1)}\otimes\cdots\otimes{\cal X}_{(n)}$ is defined by
\be
	\Delta^{(n+1)}({\cal X}) = (\Delta\otimes {\rm id}^{\otimes (n-1)})
	(\Delta^{(n)}({\cal X}))\,,
	\qquad
	\Delta^{(2)} = \Delta\,.
\ee
The representation $X$ of ${\cal X}$ on the tensor algebra 
${\cal T}({\mathscr H})=\bigoplus_{n=0}^\infty{\mathscr H}^{\otimes n}$
is the direct sum of the tensor product representations:
\be
	X=1\oplus U({\cal X}) \bigoplus_{n=2}^{\infty} U^{(n)}({\cal X})\,,
\ee
where the first part $1$ of $X$ is the trivial representation.

For the  untwisted Poincar\'e UEA,
the coproduct $\Delta_0$ is given by\footnote{
For an element ${\cal X}$ of ${\cal U}$ and an element 
${\cal Y}={\cal Y}_{(1)}\otimes{\cal Y}_{(2)}$ of ${\cal U}\otimes{\cal U}$,
we define ${\cal X}^{(n)}_i$, ${\cal X}^{(n)}$,  ${\cal Y}^{(n)}_{ij}$ and ${\cal Y}^{(n)}$
which are elements of ${\cal U}^{\otimes n}$ as
\ba
	&&{\cal X}^{(n)}_i = 1\otimes\cdots\otimes \overset{i^{\rm th}}{\cal X}
	\otimes\cdots\otimes1\,,
	\qquad
	{\cal X}^{(n)} = \sum_{i=1}^n {\cal X}^{(n)}_i\,,
	\nonumber\\
	&&{\cal Y}^{(n)}_{ij} = 1\otimes\cdots\otimes \overset{i^{\rm th}}{\cal Y}_{(1)}
	\otimes\cdots\otimes\overset{j^{\rm th}}{\cal Y}_{(2)}
	\otimes\cdots\otimes1\,,
	\qquad
	{\cal Y}^{(n)} = \sum_{1\le i<j\le n} {\cal Y}^{(n)}_{ij}\,.
\ea}
\be
	\Delta_0({\cal X})={\cal X}^{(2)}\,,
	\qquad
	\Delta_0^{(n)}({\cal X})={\cal X}^{(n)}\,,
\ee
and  the resulting tensor algebra representation $X_0$ of ${\cal X}$ is
\be
	X_0=1\oplus U({\cal X}) \bigoplus_{n=2}^{\infty} U_0^{(n)}({\cal X})\,,
	\qquad
	U_0^{(n)}({\cal X})=U^{\otimes n}(\Delta_0^{(n)}({\cal X}))\,.
\ee
In order to construct the Fock space we shall need the
flip operation $\pi_{ij}$ on ${\cal U}^{\otimes n}$ given by
\be	
	\pi_{ij}({\cal X}_1\otimes\cdots\otimes{\cal X}_n)=
	{\cal X}_1\otimes\cdots\otimes\overset{i^{\rm th}}{{\cal X}_j}
	\otimes\cdots\otimes\overset{j^{\rm th}}{{\cal X}_i}
	\otimes\cdots\otimes{\cal X}_n\, .
\ee
The n-coproduct $\Delta_0^{(n)}$ is then invariant under the flip:
\be	
	\pi_{ij}(\Delta_0^{(n)}({\cal X}))=\Delta_0^{(n)}({\cal X})\,,
\ee
and the tensor product representation $U^{(n)}({\cal X})$
commutes with the analogously defined flip maps $\Pi_{ij}$'s on ${\mathscr H}^{\otimes n}\,$:
\be	
	[\,\Pi_{ij}\,,\,U^{(n)}_0({\cal X})\,]=0\,.
\ee
Therefore, ${\mathscr H}^{\otimes n}$ is reducible and
contains two Fock spaces, the invariant eigenspaces of $\Pi_{ij}$.
Eigenvalue $+1$ (totally symmetric case) corresponds to bosons and
$-1$ (totally antisymmetric case) corresponds to fermions.
In the following, we shall concentrate on the bosonic Fock space $\mathscr F$,
which is obtained by projecting the tensor algebra ${\cal T}({\mathscr H})$
on the $+1$ eigenspace of $\Pi_{ij}\,$:
\be
	{\mathscr F}_0={\cal S}_0\,{\cal T}({\mathscr H})\,.
\ee
Here the projector ${\cal S}_0$ is the symmetrization map and
 n-boson states are thus given by
\be
	|p_1,\dots,p_n\ra_0={\cal S}_0\,|p_1\ra\otimes\cdots\otimes|p_n\ra\,,
\ee
and their scalar product  is given by
\be
	{}_0\la p_1,\dots,p_n|q_1,\dots,q_n\ra_0=
	\sum_{\mathscr P}\delta(p_1-q_{{\sst\mathscr P}1})\cdots\delta(p_1-q_{{\sst\mathscr P}n})\,,
\ee
where the sum $\sum_{\mathscr P}$ is over all permutations.

We now turn to the twisted Poincar\'e UEA which has the same algebra structure 
as the untwisted one but is equipped with a {twisted} coproduct $\Delta_\theta\,$:
\be
	\Delta_\theta({\cal X})={\cal F}\,\Delta_0({\cal X})\,{\cal F}^{-1}\,,
	\qquad
	{\cal F}=\exp{\cal G}\,,
	\qquad
	{\cal G}=\frac i2\, \theta^{\mu\nu}\, {\cal P}_\mu \otimes {\cal P}_\nu\,.
\ee
Using the notations defined in the previous footnote,
the corresponding n-coproduct $\Delta_\theta^{(n)}$ is given by
\footnote{see the appendix for a proof.}
\be
	\Delta_\theta^{(n)}({\cal X})={\cal F}_{n}\,\Delta_0^{(n)}({\cal X})\,{\cal F}_{n}^{-1}\,,
	\qquad
	{\cal F}_{n}\equiv\exp{\cal G}^{(n)}\,.
	\label{rel_cop}
\ee
Since the deformation of the Poincar\'e UEA affects only its coproduct,
the UIR space ${\mathscr H}$ and the tensor product space
${\mathscr H}^{\otimes n}$ are not changed
but the action of the algebra on ${\mathscr H}^{\otimes n}$ is deformed as
\be
	U_\theta^{(n)}({\cal X})=U^{\otimes n}(\Delta_\theta^{(n)}({\cal X}))\,.
\ee
Consequently, the tensor algebra representation $X_\theta$ of ${\cal X}$ is also deformed as
\be
	X_\theta=1\oplus U({\cal X}) \bigoplus_{n=2}^{\infty} U^{(n)}_\theta({\cal X})\,.
\ee
On the other hand, from eq.(\ref{rel_cop}) we get
\be
	U^{(n)}_\theta({\cal X})=
	U^{\otimes n}({\cal F}_{n})\,U^{(n)}_0({\cal X})\,U^{\otimes n}({\cal F}_{n})^{-1}\,,
\ee
so if we define $F$ as
\be
	F=1\oplus 1 \bigoplus_{n=2}^{\infty} U^{\otimes n}({\cal F}_{n})\,,
\ee
we obtain the similarity relation between $X_0$ and $X_\theta\,$:
\be
	X_\theta=F\,X_0\,F^{-1}\,.\label{sim}
\ee
A similar relation can be in fact established for a general counital 2-cocycle ${\cal F}$ \cite{Fiore:1996tr}.

Contrary to the undeformed case,
the actions of the twisted Poincar\'e UEA do not commute 
with the flips $\Pi_{ij}$. From the similarity relation eq.(\ref{sim}) we see
that
they commute with
 the twisted flips $F\,\Pi_{ij}\,F^{-1}\,$ 
which still have eigenvalues $\pm1$ and associated eigenspaces,
a bosonic Fock space with the eigenvalue $1$ and
a fermionic Fock space with the eigenvalue $-1\,$.
It follows that the bosonic Fock space and a twisted n-bosons state are given by
\be
	{\mathscr F}_\theta={\cal S}_\theta\,{\cal T}({\mathscr H})\,,
	\qquad
	|p_1,\dots,p_n\ra_\theta={\cal S}_\theta\,|p_1\ra\otimes\cdots\otimes|p_n\ra\,,
\ee
where the projector ${\cal S}_\theta$ is the {twisted} symmetrization map with 
respect to the twisted flip $F\,\Pi_{ij}\,F^{-1}\,$ and
is related to the undeformed symmetrization map 
by ${\cal S}_\theta=F\,{\cal S}_0\,F^{-1}\,$.

Since $F$ is diagonal when acting on  $|p_1\ra\otimes\cdots\otimes|p_n\ra$, 
we  obtain a simple relation between the twisted and the untwisted states as
\be
	F\,|p_1,\dots,p_n\ra_0=f^\theta_{p_1\cdots p_n}\,|p_1,\dots,p_n\ra_\theta\,,
	\label{rel_state}
\ee
where $f^\theta_{p_1\cdots p_n}$ is the eigenvalue of $F$ with
 eigenstate $|p_1\ra\otimes\cdots\otimes|p_n\ra\,$.
Explicitly, it is given by
\be
	f^\theta_{p_1\cdots p_n}=e^{\frac i2\sum_{i<j}p_i\theta p_j}\,,
	\qquad
	F\,|p_1\ra\otimes\cdots\otimes|p_n\ra=f^\theta_{p_1\cdots p_n}\,|p_1\ra\otimes\cdots\otimes|p_n\ra\,.
\ee

The action of the translations on the
twisted states are unchanged but
the action of the Lorentz transformations is deformed as
\be
	U_\theta(\Lambda)\,|p_1,\dots,p_n\ra_\theta=f^\theta_{\Lambda p_1\cdots \Lambda p_n}
	f^{-\theta}_{p_1\cdots p_n}\,|\Lambda p_1,\dots,\Lambda p_n\ra_\theta\,.
	\label{twist action}
\ee
It is important to notice that the symmetric states $F\,|p_1,\dots ,p_n\ra_0$ 
do not transform covariantly
under the twisted Poincar\'e transformations.
They rather transform as do the untwisted states 
under the untwisted Poincar\'e
transformations. 
If we demand that the asymptotic states of the theory
transform covariantly with $\Delta_\theta$ then they should be given by
the states $|p_1,\dots p_n\ra_{\theta}\,$. These in turn will determine the creation
and annihilation operators.

Notice from eq.(\ref{rel_state}) that the states $|p_1,\dots,p_n\ra_\theta$
are not symmetric, the exchange of  $p_i$ and $p_j$  multiplies the
state by a phase:
\be
	|p_1,\dots,\overset{i^{\rm th}}{p_j},\dots,\overset{j^{\rm th}}{p_i},\dots ,p_n\ra_\theta
	= e^{ip_i\theta p_j}\,|p_1,\dots,p_n\ra_\theta\,.
	\label{twist sym}
\ee

The determination of the creation
and annihilation operators will also 
depend on the scalar product of two  states which
is also changed by the twisting and 
is given by
\be
	{}_\theta\la p_1,\dots,p_n|q_1,\dots,q_n\ra_\theta=
	\sum_{\mathscr P} f^\theta_{p_1\dots p_n} f^{-\theta}_{q_1\dots q_n}\,
	\delta(p_1-q_{{\sst\mathscr P}1})\cdots\delta(p_1-q_{{\sst\mathscr P}n})\,.
\ee

\section{Field operators and the Moyal product}

A scalar quantum field is an operator acting on the  twisted Fock space and 
 can be expressed in terms of 
the twisted creation and annihilation operators. 
A covariant field satisfies
\be
	\Phi(\Lambda x)=U(\Lambda)\,\Phi(x)\,U^{-1}(\Lambda)\,,\label{cova}
\ee
where $\Lambda$ is a Poincar\'e transformation.
In the untwisted case, the covariance relation (\ref{cova}) determines
completely
a scalar field linear in the creation and annihilation operators (see, for
instance, Section 2 of {\cite{Joung:2006gj} for a short proof).
The resulting scalar field reads
\be
	\Phi_0(x)=\phi_0^{(+)}(x)+\phi_0^{(-)}(x)\,,
	\qquad
	\phi_0^{(+)}(x)=\int d\mu(p)\,e^{ip\cdot x}\,a_0^\dagger(p)\,,
\ee
where $\phi^{(-)}$ is the  hermitian conjugate of $\phi^{(+)}$ and
$d\mu(p)= d^{d-1}p\,(2\pi)^{1-d \over 2} (2\omega(p))^{-\frac12}$ is the invariant
measure.
The creation operators are defined by their action on the multi-particle
states as
\be
	a_0^\dagger(p)\,|p_1,\dots,p_n\ra_0=|p,p_1,\dots,p_n\ra_0\,.
\ee
From the symmetry and the scalar product, we obtain
the commutation relations:
\be
	[\,a_0^\dagger(p)\,,\,a_0^\dagger(q)\,]=[\,a_0(p)\,,\,a_0(q)\,]=0\,,
	\qquad
	[\,a_0(p)\,,\,a_0^\dagger(q)\,]=\delta(p-q)\,.
\ee

We now define the twisted creation and annihilation operators 
in the same way by
\be
	a_\theta^\dagger(p)\,|p_1,\dots,p_n\ra_\theta=
	|p,p_1,\dots,p_n\ra_\theta\,,
\ee
and also from the symmetry and the scalar product of the 
twisted states we obtain the following
deformed version of the commutation relations:
\ba
	&&a_\theta^\dagger(p)\,a_\theta^\dagger(q)=e^{-ip\theta q}\,a_\theta^\dagger(q)\,a_\theta^\dagger(p)\,,
	\qquad
	a_\theta(p)\,a_\theta(q)=e^{-ip\theta q}\,a_\theta(q)\,a_\theta(p)\,,
	\nonumber\\
	&&\qquad\qquad
	a_\theta(p)\,a_\theta^\dagger(q)=e^{ip\theta q}\,a_\theta^\dagger(q)\,a_\theta(p)+\delta(p-q)\,.
	\label{cr aF}
\ea
The relation between the twisted and untwisted states (\ref{rel_state})
can be written in terms of the above defined creation and annihilation operators:
\ba
	f^\theta_{p_1\cdots p_n}\,|p_1,\dots,p_n\ra_\theta &=&
	f^\theta_{p_1\cdots p_n}\,a_\theta^\dagger(p_1)\cdots a_\theta^\dagger(p_n)|\Omega\ra
	=\left(e^{\frac i2 p_1\theta P}\,a_\theta^\dagger(p_1)\right)\cdots
	\left(e^{\frac i2 p_n\theta P}\,a_\theta^\dagger(p_n)\right)|\Omega\ra
	\nonumber\\
	&=&F\,|p_1,\dots,p_n\ra_0=
	\left(F\,a_0^\dagger(p_1)\,F^{-1}\right)\cdots\left(F\,a_0^\dagger(p_1)\,F^{-1}\right)|\Omega\ra\,.
\ea
where $P^\mu=P_\theta^\mu=P_0^\mu$ is the Fock 
representation of the momentum generator ${\cal P}^\mu\,$.
From the above equation, we obtain a simple relation between the
twisted and untwisted creation operators as
\be
	F\,a_0^\dagger(p)\,F^{-1}=e^{\frac i2 p\theta P}\,a_\theta^\dagger(p)\,.
	\label{rel cao}
\ee
From the unitary transformation of untwisted creation 
operator by a finite Lorentz transformation $\Lambda$:
\be
	U_0(\Lambda)\,a_0^\dagger(p)\,U_0(\Lambda)^{-1}=a_0^\dagger(\Lambda p)\,,
\ee
we obtain the unitary transformation of its twisted counterpart as 
\ba
	&&U_\theta(\Lambda)\left( e^{\frac i2 p\theta P}\,a_\theta^\dagger(p)\right) U_\theta(\Lambda)^{-1}
	=e^{\frac i2 (\Lambda p)\theta P}\,a_\theta^\dagger(\Lambda p)\,,
	\nonumber\\
	&&U_\theta(\Lambda)\,a_\theta^\dagger(p)\,U_\theta(\Lambda)^{-1}
	=e^{\frac i2(\Lambda p)\theta P-\frac i2 p\theta (\Lambda P)}\,a_\theta^\dagger(\Lambda p)\,.
	\label{aF}
\ea

At this stage, we can compare the above defined n-boson states and 
creation and annihilation operators
with those of previous works. 
The twisted n-particle state 
 considered by Balachandran et al \cite{Balachandran:2006pi}
 have the same statistics as  our states $|p_1,\cdots,p_n\ra_\theta\,$;
 the relation between the creation operators $a_0^\dagger$ and $a_\theta^\dagger$
 differs however by a unitary transformation $F$.
 The states considered by Fiore and Schupp \cite{Fiore:1996tr},
are totally symmetric and correspond  to
$F\,\,|p_1,\cdots,p_n\ra_0\,$.
The relation between the twisted and untwisted 
creation and annihilation operators is obtained
by Fiore \cite{Fiore:1996pm} for general twisted Lie algebra.

Having determined the creation and annihilation operators we can now
turn to the construction of the field operator. The scalar 
field which was used in previous studies \cite{Balachandran:2005eb,Fiore:2007vg}
is given by
\be
	\Phi_{\rm nc\, \theta}(x)=\phi_{\rm nc\, \theta}^{(+)}(x)+\phi_{\rm nc\, \theta}^{(-)}(x)\,,
	\qquad
	\phi_{\rm nc\, \theta}^{(+)}(x)=\int d\mu(p)\,e^{ip\cdot
	x}\,a_\theta^\dagger(p)\,.
	\ee
and is obtained from $\Phi_0$ by the replacement of $a_0^\dagger$ by $a_\theta^\dagger$.
 The so defined field $\Phi_{\rm nc\, \theta}(x)$ is however not covariant.
Using eq.(\ref{aF}), we deduce the action of a  Lorentz transformation:
\ba
	U_\theta(\Lambda)\,\Phi_{\rm nc\, \theta}(x)\,U_\theta(\Lambda)^{-1}
	&=&\int d\mu(p)\,e^{ip\cdot (\Lambda x)}\,e^{\frac i2 p\theta P-\frac i2 (\Lambda^{-1}p)\theta (\Lambda P)}
	\,a_\theta^\dagger(p)
	\nonumber\\
	&=&\Phi_{\rm nc\, \theta}(\Lambda x+{\st\frac12}
	(\theta-\Lambda\theta\Lambda)P)\,.
	\label{noncov field}
\ea
In fact, we can see from eq.(\ref{aF}) that
there is no covariant field which is linear in the 
creation and annihilation operators.
 If we loosen up the linearity condition, we can get one. In order to see this, 
 let us first define  $\tilde{a}_\theta^\dagger(p)$ by
 \be
	\tilde a_{\theta}^\dagger(p)=e^{\frac i2 p\theta P}\,a_\theta^\dagger(p)
	=e^{\frac i2\int d\mu(q)\,p
	\theta q\,a_\theta^\dagger(q)\,a_\theta(q)}a_\theta^\dagger(p)\,.
	\label{cov creat}
\ee
 From the transformation law (\ref{aF}), we have
 \be
 U_\theta(\Lambda)\,\tilde a^\dagger_{\theta}(p)U_\theta(\Lambda)^{-1}=
 \tilde a^\dagger_{\theta}(\Lambda p)\,,
 \ee
 and  from eq.(\ref{rel cao}), it satisfies the usual commutation relations:
\be
	[\,\tilde a_{\theta}^\dagger(p)\,,\,\tilde a_{\theta}^\dagger(q)\,]=[\,\tilde a_{\theta}(p)\,,\,\tilde a_{\theta}(q)\,]=0\,,
	\qquad
	[\,\tilde a_{\theta}(p)\,,\,\tilde a_{\theta}^\dagger(q)\,]=\delta(p-q)\,.
	\label{cr acF}
\ee
The states generated by successive applications of $\tilde a_{\theta}^\dagger(p)$ on $|\Omega\ra$
are $F\,|p_1,\dots,p_n\ra_0\,$.
We deduce that the  field operator given by
\be
	\Phi_\theta(x)=\phi_\theta^{(+)}(x)+\phi_\theta^{(-)}(x)\,,
	\qquad
	\phi_\theta^{(+)}(x)=\int d\mu(p)\,e^{ip\cdot x}\,\tilde a_\theta^\dagger(p)\,.
	\label{cF}
\ee
is covariant under the Poincar\'e transformations.
Furthermore, from eq.(\ref{rel cao}) it is related to the untwisted one 
by a unitary  transformation:
\be
	F\,\Phi_0(x)\,F^{-1}=\Phi_{\theta}(x)\,.
	\label{equiv rel field}
\ee
Therefore, since $F$ leaves the vacuum state invariant,
the QFT with the covariant field $\Phi_\theta$
(\ref{cF}) based on twisted Poincar\'e UEA leads to the same n-point 
functions as the untwisted one.
This covariant field $\Phi_\theta$,
 which was not considered in the
 literature before, allows to
construct local interaction Hamiltonian's which are manifestly invariant
under the twisted symmetry. For example the interaction action $-\int
V(\Phi_\theta)$ with $V$ an arbitrary potential,
is manifestly invariant under the Poincar\'e transformations.
Notice that the Moyal product is not needed to achieve the invariance under the
twisted Poincar\'e symmetry when using the covariant field $\Phi_\theta\,$.

We now turn to examine the relation with the Moyal product 
which was at the
origin of this twisted symmetry.
To cover the most general case, we shall
consider a generalized version of the Moyal product, which gives
 the $*$-product between two fields at different spacetime points, 
 with an arbitrary
twist parameter $\vartheta\,$ as
\be
	(f*_{\sst\vartheta}g)(x,y)=e^{\frac i2 \partial_x\vartheta\partial_y}f(x)g(y)\,.
\ee
The $\vartheta=0$ case corresponds to the usual pointwise product.
The $*_{\sst\vartheta}$-product of n plane waves is given by
\be
	e^{ip_1\cdot x_1}*_{\sst\vartheta}\cdots*_{\sst\vartheta}e^{ip_n\cdot x_n}
	=f^{-\vartheta}_{p_1\cdots p_n}\,
	e^{ip_1\cdot x_1}\cdots e^{ip_n\cdot x_n}\,.
	\label{star wave}
\ee
Consider now the field product 
\be
	(\Phi_{\rm  \theta}*_{\sst\vartheta}
	\cdots*_{\sst\vartheta}\Phi_{\rm  \theta})(x_1,\cdots,x_n)\,.
	\label{pr}
\ee
It can be   written as 
$\sum_{s_i=\pm}(\phi_{\rm  \theta}^{(s_1)}*_{\sst\vartheta}
\cdots*_{\sst\vartheta}\phi_{\rm  \theta}^{(s_n)})(x_1,\cdots,x_n)$
with
\ba
	&&
	(\phi_{\rm  \theta}^{(s_1)}*_{\sst\vartheta}\cdots *_{\sst\vartheta}
	\phi_{\rm  \theta}^{(s_n)})(x_1,\cdots,x_n)=
	\nonumber\\&&=
	\int d\mu(p_1)\cdots d\mu(p_n)\,
	e^{s_1 ip_1\cdot x_1}*_{\sst\vartheta}\cdots *_{\sst\vartheta}e^{s_n ip_n\cdot x_n}\,
	\tilde a_\theta^{(s_1)}(p_1)\cdots \tilde a_\theta^{(s_n)}(p_n)\,,
	\label{equiv}
\ea
where $\tilde a^{(+)}_\theta=\tilde a^\dagger_\theta$ and 
$\tilde a^{(-)}_\theta=\tilde a_\theta\,$.
Using eq.(\ref{star wave}) and eq.(\ref{cov creat}), the integrand can be rewritten as
\ba
	&&
	e^{s_1 ip_1\cdot x_1}*_{\sst\vartheta}\cdots *_{\sst\vartheta}e^{s_n ip_n\cdot x_n}\,
	\tilde a_\theta^{(s_1)}(p_1)\cdots \tilde a_\theta^{(s_n)}(p_n)
	\nonumber\\
	&&=
	e^{s_1 ip_1\cdot x_1}\cdots e^{s_n ip_n\cdot x_n}\,
	f^{-\vartheta}_{s_1p_1\cdots s_np_n}
	\left(e^{\frac i2 s_1p_1\theta P}\, a_{\theta}^{(s_1)}(p_1)\,\right)\cdots
	\left(e^{\frac i2 s_np_n\theta P}\, a_{\theta}^{(s_n)}(p_n)\,\right)
	\nonumber\\
	&&=
	e^{s_1 ip_1\cdot x_1}\cdots e^{s_n ip_n\cdot x_n}\,f^{-\vartheta+\theta}_{s_1p_1\cdots s_np_n}\,
	 a_{\theta}^{(s_1)}(p_1)\cdots  a_{\theta}^{(s_n)}(p_n)\,
	e^{\frac i2(s_1p_1+\cdots+s_np_n)\theta P}
	\nonumber\\
	&&=
	e^{s_1 ip_1\cdot x_1}*_{\sst\vartheta-\theta}
	\cdots*_{\sst\vartheta-\theta}e^{s_n ip_n\cdot x_n}\,
	 a_{\theta}^{(s_1)}(p_1)\cdots  a_{\theta}^{(s_n)}(p_n)\,
	e^{\frac i2(s_1p_1+\cdots+s_np_n)\theta P}\,.
	\label{pro}
\ea
Taking $x_1=x_2=\dots=x_n=x$ in eq.(\ref{pr}), 
using eq.(\ref{pro}) and integrating over $x$
gives 
\be
	\int dx\ \Phi_{\theta}*_{\sst\vartheta}
	\cdots *_{\sst\vartheta}\Phi_{\theta}
	(x)\,=\,
	\int dx\ \Phi_{\rm nc \theta}*_{\sst\vartheta-\theta}
	\cdots *_{\sst\vartheta-\theta}\Phi_{\rm nc \theta}(x)\,.
\ee
From the similarity relation (\ref{equiv rel field}), we can  also cast the
above integral in the form
\be
	F \left(\int dx\ \Phi_{0}*_{\sst\vartheta}
	\cdots *_{\sst\vartheta}\Phi_{0}(x) \right) F^{-1}\,.
\ee
This shows that the invariant interaction given by a monomial potential 
$\int dx\,\Phi_{\theta}^n(x)$ can also be written as
$\int dx\ \Phi_{\rm nc \theta}*_{\sst-\theta}
\cdots *_{\sst-\theta}\Phi_{\rm nc \theta}(x)$ or as
$F\left(\int dx\,\Phi_{0}^n(x)\right)F^{-1}$.

Acting with the field product (\ref{pr}) on the vacuum state, 
the last factor in eq.(\ref{pro})  drops out and we get 
\ba
(\Phi_{\theta}*_{\sst\vartheta}\cdots *_{\sst\vartheta} 
\Phi_{\theta})(x_1,\cdots,x_n)
	|\Omega\ra &=&
	(\Phi_{\rm nc \theta}*_{\sst\vartheta-\theta}
	\cdots *_{\sst\vartheta-\theta}\Phi_{\rm nc \theta})(x_1,\cdots,x_n)\,
	|\Omega\ra
	\nonumber\\
	&=&
	F\,(\Phi_{0}*_{\sst\vartheta}\cdots *_{\sst\vartheta} \Phi_{0})(x_1,\cdots,x_n)\,
	|\Omega\ra\,,
\ea
where we used eq.(\ref{equiv rel field}) to get the last equality. 
This shows that the n-point functions with different 
choices of fields and products are related to each other
in a simple way.
 The case $\vartheta=\theta$  shows that using $\Phi_0$ with the Moyal product, 
which defines NCFT in the Moyal plane,
is equivalent to using the non-covariant field $\Phi_{\rm nc\theta}$ with the
 usual product and is not invariant under the twisted symmetry.
 The case $\vartheta=0$ leads to covariant results and 
 shows that using $\Phi_\theta$ with the usual product 
 is equivalent to $\Phi_{\rm nc\theta}$ with a 
 twisted product but is also related to
 the untwisted $\Phi_0$ with the usual product. 
 This agrees with the conclusions
 of  \cite{Balachandran:2005pn,Abe:2006ig,Fiore:2007vg}.

The theory is fully determined by the asymptotic states and the $S$-matrix.
We argued that the asymptotic states which transform covariantly under the
twisted symmetry are $|p_1,\cdots,p_n\ra_\theta=f_{p_1\cdots p_n}^{-\theta} F\,|p_1,\cdots, p_n \ra_0\,$.
On the other hand the $S$-matrix should commute with the transformations
$U_\theta({\cal X})\,$. This can be easily realized with  
an interaction Hamiltonian local in $\Phi_\theta$.
Since $\Phi_\theta=F\,\Phi_0\,F^{-1}$, the obtained $S$-matrix is unitarily
equivalent to the corresponding untwisted $S$-matrix $S_0$ :
$S=F\,S_0\,F^{-1}$.
The resulting matrix elements are related by phases as
\be
	_\theta\la q_1,\dots, q_n|\,S\,|p_1,\dots, p_m\ra_\theta=
	f_{q_1,\dots, q_n}^{\theta}f_{p_1,\dots, p_m}^{-\theta}\,
	_0\la q_1,\dots, q_n|\,S_0\,|p_1,\dots, p_m\ra_0\,.
\ee
This is the main consequence of the invariance under the twisted Poincar\'e
symmetry.

Our conclusions apply to the twist given by eq.(\ref{Moyal twist}), other twists
of the Poincar\'e are also possible \cite{Lukierski:2005fc}, 
it would be interesting to explore 
the consequences of the invariance under these twisted symmetries.

\acknowledgments

We would like to thank John Madore, Karim Noui and Renaud Parentani
for many useful discussions.

\appendix

\section{Twisted n-coproduct}

In this Appendix we prove eq.(\ref{rel_cop}) by induction.
First, we have
\be
	\Delta_\theta^{(2)}({\cal X})=\Delta_\theta({\cal X})={\cal F}\,\Delta_0({\cal X})\,{\cal F}^{-1}
	={\cal F}_{2}\,\Delta_0^{(2)}({\cal X})\,{\cal F}_{2}^{-1}\,.
\ee
Assuming $\Delta_\theta^{(n)}({\cal X})={\cal F}_{n}\,\Delta_0^{(n)}({\cal X})\,{\cal F}_{n}^{-1}\,$,
the next order can be calculated as
\ba
	\Delta_\theta^{(n+1)}({\cal X})&=&(\Delta_\theta\otimes {\rm id}^{\otimes (n-1)})(\Delta_\theta^{(n)}({\cal X}))
	\\
	&=&e^{{\cal G}^{(n+1)}_{12}}
	(\Delta_0\otimes {\rm id}^{\otimes (n-1)})	({\cal F}_{n}\,\Delta_0^{(n)}({\cal X})\,{\cal F}_{n}^{-1})\,
	e^{-{\cal G}^{(n+1)}_{12}}
	\nonumber\\
	&=&e^{{\cal G}^{(n+1)}_{12}}
	(\Delta_0\otimes {\rm id}^{\otimes (n-1)})({\cal F}_{n})
	\,\Delta_0^{(n+1)}({\cal X})\,
	(\Delta_0\otimes {\rm id}^{\otimes (n-1)})({\cal F}_{n}^{-1})\,
	e^{-{\cal G}^{(n+1)}_{12}}\,.
	\nonumber
\ea
Since $\Delta_0\otimes {\rm id}^{\otimes (n-1)}$ is linear, we get
\ba	
	(\Delta_0\otimes {\rm id}^{\otimes (n-1)})({\cal F}_{n})
	&=&(\Delta_0\otimes {\rm id}^{\otimes (n-1)})
	\left(\exp\left[{\st\sum}_{1\le i<j\le n}\,{\cal G}^{(n)}_{ij}\right]\right)
	\nonumber\\
	&=&\exp\left[{\st\sum}_{1\le i<j\le n}\,(\Delta_0\otimes {\rm id}^{\otimes (n-1)})({\cal G}^{(n)}_{ij})\right]
	\nonumber\\
	&=&\exp\left[{\st\sum}_{2\le j\le n}\,(\Delta_0\otimes {\rm id}^{\otimes (n-1)})({\cal G}^{(n)}_{1j})
	+{\st\sum}_{3\le i<j\le n+1}\,{\cal G}^{(n+1)}_{ij}\right]
	\nonumber\\
	&=&\exp\left[{\st\sum}_{3\le j\le n}\,({\cal G}^{(n+1)}_{1j}+{\cal G}^{(n+1)}_{2j})
	+{\st\sum}_{3\le i<j\le n+1}\,{\cal G}^{(n+1)}_{ij}\right].
\ea
Finally we obtain
\be
	e^{{\cal G}^{(n+1)}_{12}}(\Delta_0\otimes {\rm id}^{\otimes (n-1)})({\cal F}_{n})
	=\exp\left[{\st\sum}_{1\le i<j\le n+1}\,{\cal G}^{(n+1)}_{ij}\right]
	={\cal F}_{n+1}\,,
\ee
and $\Delta_\theta^{(n+1)}({\cal X})={\cal F}_{n+1}\,\Delta_0^{(n+1)}({\cal X})\,{\cal F}_{n+1}^{-1}\,$.

\bibliographystyle{JHEP.bst}

\bibliography{ref}

\end{document}